\documentclass[article,nojss]{jss}
\usepackage{amsmath}
\usepackage{amssymb}
\usepackage{graphicx}
\usepackage{mathtools}
\usepackage{tabularx}
\usepackage{algorithm}
\usepackage{algpseudocode}
\usepackage{rotating}
\usepackage{booktabs}

\author{Samuel Watson\\University of Birmingham}
\title{Twenty ways to estimate the Log Gaussian Cox Process model with point and aggregated case data: the \pkg{rts2} package for \proglang{R}}

\Plainauthor{Samuel Watson} 
\Plaintitle{Twenty ways to estimate the Log Gaussian Cox Process model with point and aggregated case data: the rts2 package for R} 
\Shorttitle{20 methods for the LGCP model} 

\Abstract{
   The \proglang{R} package \pkg{rts2} provides data manipulation and model fitting tools for Log Gaussian Cox Process (LGCP) models. LGCP models are a key method for disease and other types of surveillance, and provide a means of predicting risk across an area of interest based on spatially-referenced and time-stamped case data. However, these models can be difficult to specify and computationally demanding to estimate. For many surveillance scenarios we require results in near real-time using routinely available data to guide and direct policy responses, or due to limited availability of computational resources. There are limited software implementations available for this real-time context with reliable predictions and quantification of uncertainty. The \pkg{rts2} package provides a range of modern Gaussian process approximations and model fitting methods to fit the LGCP, including estimation of covariance parameters, using both Bayesian and stochastic Maximum Likelihood methods. The package provides a suite of data manipulation tools. We also provide a novel implementation to estimate the LGCP when case data are aggregated to an irregular grid such as census tract areas.
}
\Plainkeywords{Markov Chain Monte Carlo Maximum Likelihood, Generalised Linear Mixed Model, R, C++, Bayesian, Gaussian Process} 

\Address{
  Samuel Watson\\
  Institute for Applied Health Research\\
  University of Birmingham \\
  Birmingham, UK\\
  E-mail: \email{S.I.Watson@bham.ac.uk}
}

\begin{document}


\section{Introduction}

The statistical analysis of geolocated and possibly time-stamped data representing cases of a phenomenon of interest can help identify areas with a high probability of high or growing risk. In this article, we focus primarily of disease surveillance, but other applications include social policy, crime, ecological, geological, or other point process phenomena. In the context of disease surveillance, prevalence mapping surveys are a rigorous method for measuring disease risk and conduct a 0/1 tests on a sample from a population in a given area to predict the spatial distribution of risk \citep{Diggle2016}. Recent examples include the mapping of seroprevalence during the SARS-CoV-2 pandemic (e.g. \citet{Riley2020}). However, such studies require significant time and resources. In many contexts there is a need for more rapid analyses of case data to be able to quickly identify and respond to emerging disease outbreaks. Again, SARS-CoV-2 provides a good example \citep{Watson2021a}, but in many low- and middle-income country settings identifying emerging clusters of diseases like cholera, malaria, or influenza can facilitate targeted intervention to prevent the development of an epidemic (e.g. \citet{Ratnayake2020a,Ratnayake2020}). Beyond emergency applications, reliable quantification of disease risk across an area of interest using routine data provides a useful tool for public health surveillance and research. Obtaining predictions in a reasonable amount of time without using excessive computational resources can facilitate their adoption.

Many healthcare and public health systems collect finely spatially-resolved and time stamped data that can be used to monitor disease epidemiology. These data may include daily hospital admissions, positive tests, or calls to public health telephone services \citep{Diggle2005}. Case data are often also provided as counts at a spatially aggregated level, such as census tracts. Geospatial statistical models provide a principled basis for generating predictions of disease epidemiology and its spatial and temporal distribution from these data. 

The \pkg{rts2} package provides a data manipulation and model fitting methods for when these data can be assumed to be a realisation from a spatio-temporal point process over the area of interest. A point, such as the residential address of a hospital admission on a particular date, is assumed to be more likely to be observed in areas of higher risk than lower risk. Analysis of these data can provide probabilistic predictions about the underlying disease risk and be used to quantify our uncertainty about the presence of ``hotspots''. 

A growing use case for spatio-temporal point process models is ``real-time'' disease surveillance that uses a stream of real-time data, such as hospital admissions, to provide rapid feedback to public health authorities. However, geospatial models can be computationally intensive and slow to run, and they can scale poorly with the size of the data. `Fast' methods like kernel-based mapping requires often difficult selection of bandwidth parameters and may have less reliable quantification of uncertainty. \pkg{rts2} provides a range of Gaussian process approximations and approximate inference, including both Bayesian and Frequest approaches, to support faster and more scalable analyses for real-time scenarios. We first describe the statistical models and model fitting and other relevant packages for R, and then provide a discussion of the use of the package and generation and interpretation of outputs.

\section{The Log Gaussian Cox Process}
Our statistical model is a Log Gaussian cox process (LGCP) \citep{Diggle2013}, whose realisation is observed on the Cartesian area of interest $A \subset \mathbb{R}^2$ and time period of interest $T \subset \mathbb{N}$ (we assume discrete time periods). The resulting data are relaisations of an inhomogeneous Poisson process with stochastic intensity function $\{\lambda(s,t):s\in A, t\in T\}$. The number of cases occuring in any locally finite random set $S \subseteq A$ is Poisson distributed conditional on $\lambda(s,t)$:
\begin{equation}
Y(S,t) \sim \text{Poisson} \left( \int_S \lambda(s,t)ds \right)
\end{equation}
The intensity is decomposed as the log-linear model:
\begin{equation}
\lambda(s,t) = r(s,t)\text{exp}(X(s,t)'\gamma + Z(s,t))
\end{equation}
where $r(s,t)$ is a spatially or spatio-temporally varying Poisson offset, typically the population density. $X(s,t)$ is a length $Q$ vector of spatially, temporally, or spatio-temporally varying covariates including an intercept and $\{ Z(s,t): s\in A, t\in T \}$ is a latent field. 

We use an auto-regressive specification for the latent field:
\begin{align}
\begin{split}
\label{eq:z}
 Z(s,1) =& (1+\rho^2)^{-1}\mathcal{Z}_1(s)\\
 Z(s,t) =& \rho Z(s,t-1) + \mathcal{Z}_t(s) \text{   for } t>1
 \end{split}
\end{align}
where $\rho$ is the auto-regressive parameter. This specification, sometimes referred to as a ``spatial innovation'' model, facilitates computationally efficient model fitting as we describe in the subsequent section. We include both Bayesian and maximum likelihood model fitting approaches in the package and, depending on the paradigm, the spatial innovation term $\{ \mathcal{Z}_t(s): s\in A \}$ is given a Gaussian process prior distribution, or is itself a Gaussian process. In both cases, the realisations of $\mathcal{Z}_t(s)$, which we notate as $\mathbf{z}$ are multivariate Gaussian distributed with mean zero and covariance $\Sigma$. We use a minimally parameterised covariance function for the Gaussian process to define the elements of $\Sigma$:
\begin{equation}
\text{cov}(\mathcal{Z}_t(s),\mathcal{Z}_t(s')) = \sigma^2 h(||s-s'||; \phi)
\end{equation}
where $||.||$ is the L2-norm. The package includes, for example, the squared exponential covariance function
\begin{equation}
h(||s-s'||;\phi) = \text{exp}\left( \frac{-(||s-s'||)^2}{\phi^2} \right)
\end{equation}
and the exponential covariance function
\begin{equation}
h(||s-s'||;\phi) = \text{exp}\left( \frac{-||s-s'||}{\phi} \right)
\end{equation}
where $\phi$ is the spatial range (or length scale) parameter.

\section{Model Fitting and Approximation}
The LGCP is complex to estimate as the likelihood does not have a closed form solution. In this section, we discuss model approximations and fitting methods. We include methods and approximations where there do not currently exist implementations in other software packages, which are discussed in Section \ref{sec:software}.

For both types of approach we use the standard practice of dividing the area of interest into a regular grid \citep{Diggle2013,Taylor2013,Taylor2015} with grid cells $S_i:i = 1,...,n$. The size of the grid cells should permit the assumption that the latent field $Z$ is approximately constant within each cell. The counts in each of the $n$ grid cells comprise our data $\mathcal{D} = \{Y(S_i,t):i=1,..,n;t=1,...,T\}$ such that our model is:
\begin{align}
\begin{split}
\label{eq:fullmodel}
    Y(S_i,t) &\sim \text{Poisson}(\lambda(S_i,t)) \\
    \lambda(S_i,t) &= r(S_i,t)\text{exp}(X(S_i,t)\gamma + Z(S_i,t)) 
    \end{split}
\end{align}
where $Z$ is as specified in Equation (\ref{eq:z}) with $\mathcal{Z}(S_i,t) \sim \text{GP}_t(\mathbf{0},h(d;\phi))$. For the Bayesian approaches we also require specification of priors and hyperpriors for the remaining model parameters as we discuss below. 

For a given data set $\mathcal{D}$, we write the $nT$-length vector of realisations of the Gaussian process model as $\mathbf{z}$. We can then write the likelihood of the model parameters as
\begin{equation}
\label{eq:lgcplik}
    L(\gamma,\theta \vert \mathcal{D}) = \int \prod_{i=1}^n \prod_{t=1}^T f_{Y \vert z}(Y(S_i,t) \vert \gamma,\mathbf{z}) dF_{z\vert \theta }(\mathbf{z}\vert \theta)
\end{equation}
where $f_{Y\vert z}$ is the Poisson probability density function with intensity given in (\ref{eq:fullmodel}) and $f_{z\vert \theta}$ is the multivariate Gaussian density with parameters $\theta = [ \sigma^2, \phi, \rho ]$. The components of the log-likelihood are therefore:
\begin{align}
\begin{split}
\label{eq:mvnll}
    \log(f_{Y \vert z}(Y(S_i,t) \vert \gamma,\mathbf{z})) &= Y(S_i,t)\eta(S_i,t) - \lambda(S_i,t) - \log(Y(S_i,t)!) \\
    \log(f_{Z \vert \theta }(\mathbf{z} \vert \theta)) &= -\frac{m}{2}\log{(2\pi)} - \frac{1}{2}\log(|\Sigma|) - \frac{1}{2}\mathbf{z}^T \Sigma^{-1} \mathbf{z}
\end{split}
\end{align}
where $\eta(S_i,t) = \log(r(S_i,t)) + X(S_i,t)\gamma + Z(S_i,t)$.

The autoregressive specification of the latent field means we can partially reduce the complexity of the log likelihood. We can write:
\begin{equation*}
    \Sigma = P \otimes \Sigma_0
\end{equation*}
where $\Sigma_0$ is the covariance matrix for a single time period and 
\begin{equation*}
    P = \begin{bmatrix}
    1 & \rho & \rho^2 & \hdots & \rho^T \\
    \rho & 1 & \rho & \hdots & \rho^{T-1} \\
    \vdots & & \ddots & \vdots & \vdots \\
    \rho^T & \rho^{T-1} & \rho^{T-2} & \hdots & 1 
    \end{bmatrix}
\end{equation*}
Letting $\mathbf{z}_{[t]}$ refer to the elements of $\mathbf{z}$ for time period $t$ and $\mathbf{v}$ be a $n \times T$ matrix formed by stacking the elements of $\mathbf{z}$ into columns with column $t$ equal to $\mathbf{z}_{[t]}$, we can then write:
\begin{align*}
    \mathbf{z}^T \Sigma^{-1} \mathbf{z} &= \mathbf{z}^T (P \otimes \Sigma_0)^{-1} \mathbf{z}\\
    &= \mathbf{z}^T \text{Vec}(\Sigma_0^{-1}\mathbf{v}P^{-1}) \\
    &= \sum_t \mathbf{z}_{[t]} \Sigma_0^{-1} \Tilde{\mathbf{v}}_{[t]}
\end{align*}
where $\Tilde{\mathbf{v}} = \mathbf{v}P^{-1}$. Similarly $\log{(\vert \Sigma \vert)} = n\log{(\vert P \vert)} + T\log{(\vert \Sigma_0 \vert)}$ Computational time with this approach is therefore linear in the number of time periods. However, evaluation of $\Sigma_0^{-1}$ is still computationally demanding and scales as $O(n^3)$, which may still be prohibitive for many real-time surveillance applications. 

\subsection{Gaussian Process Approximations}
Evaluating the likelihood is computationally expensive owing to the need to invert $\Sigma$. Computational complexity and memory requirements to evaluate the log likelihood in Equation (\ref{eq:mvnll}) scale as $O(Tn^3)$ and $O(Tn^2)$, respectively. These models therefore quickly become infeasible with even moderately sized data sets, which can limit their use. We offer an option to skip model fitting with ``known'' covariance parameter values, which can be seen as akin to the problem of choosing a bandwidth for kernel-based approaches. We provide a function to estimate the empirical semivariogram to support parameter choice. However, for fitting procedures including the covariance parameters, there are two main classes of approximation to the likelihood of a Gaussian process we include. 

\subsubsection{Nearest Neighbour Gaussian Process}
The multivariate Gaussian likelihood can be rewritten as the product of the conditional densities:
\begin{equation*}
    f(z) = f(z_1) \prod_{i=2}^n f(z_i|z_1,...,z_{i-1})
\end{equation*}
\citet{Vecchia1988} proposed that one can approximate $f(z)$ by limiting the conditioning sets for each $z_j$ to a maximum size of $m$. For geospatial applications, \citep{Datta2016,Datta2016b} proposed the nearest neighbour Gaussian process (NNGP), in which the conditioning sets are limited to the $m$ `nearest neighbours' of each observation. 

Let $\mathcal{N}_j$ be the set of up to $m$ nearest neighbours of $j$ with index less than $j$. We use the notation $\Sigma_{\mathcal{N},\mathcal{M}}$ to represent the submatrix of $\Sigma$ with rows in $\mathcal{N}$ and columns in $\mathcal{M}$, where these sets could be of size one. The approximation is:
\begin{equation*}
    f(z) \approx f(z_1) \prod_{i=2}^n f(z_i|\mathcal{N}_i)
\end{equation*}
which leads to:
\begin{align}
\begin{split}
\label{eq:za}
    z_1 &= \eta_1 \\
    z_j &= \sum_{i=1}^p a_{ji}z_{\mathcal{N}_{ji}}
    \end{split}
\end{align}
where $\mathcal{N}_{ji}$ is the $i$th nearest neighbour of $j$.

Equation (\ref{eq:za}) can be more compactly written as $z = Az + \eta$ where $A$ is a sparse, strictly lower triangular matrix, $\eta \sim N(0,D)$ with $D$ a diagonal matrix with entries $D_{11} = \text{Var}(z_1)$ and $D_{ii} = \text{Var}(w_i | \mathcal{N}_i)$ \citep{Finley2019}. The approximate covariance matrix can then be written as $\Sigma_0 \approx (I-A)^{-1}D(I-A)^{-T}$ such that the quadratic form is approximated as:
\begin{equation*}
    \sum_t \mathbf{z}_{[t]} (I-A)^T D^{-1}(I-A) \Tilde{\mathbf{v}}_{[t]}
\end{equation*}
As $I-A$ is upper triangular, the quadratic form can be calculated cheaply using forward substitution, iterating only over the nearest neighbours for each row. The log determinant is: 
\begin{equation*}
    \log(|\Sigma_0|) \approx \sum_{i=1}^m -\log(D_{ii}) 
\end{equation*}

\citet{Finley2019} provide algorithms for the calculation of the matrices $A$ and $D$. The non-zero elements of $A$ are given by the linear system $\Sigma_{0;\mathcal{N}_j,\mathcal{N}_j}A_{j,\mathcal{N}_j}^T = \Sigma_{0;\mathcal{N}_j,j}$ and the diagonal elements of matrix $D$ are $D_{j,j} = \Sigma_{0;j,j} -  A_{j,\mathcal{N}_j}\Sigma_{0;\mathcal{N}_j,j}$. The NNGP has computational complexity scaling with $O(Tnm^3)$ and so the log-likelhood is now linear in both time and grid size.


 There are two main factors that can affect the accuracy of the NNGP approximation. The ordering of the grid cells in terms of their indices can affect performance of the approximation. \citet{Guinness2018} compares the performance of several different ordering schemes. They suggest that a ``minimax'' scheme, in which the next observation in the order is the one which maximises the minimum distance to the previous observations, often performed best, although in several scenarios ordering in terms of the vertical coordinate, or at random provided comparable performance. The number of neighbours did not appear to affect the relative performance of the orderings; they considered both 30 and 60, although \citet{Datta2016} uses 15 nearest neighbours in their simulations.

\subsubsection{Hilbert Space Gaussian Process}
Low, or reduced, rank approximations aim to approximate the matrix $\Sigma$ with a matrix $\Tilde{\Sigma}$ with rank $m < n$. The optimal low-rank approximation is $\Tilde{\Sigma} = \Phi \Lambda \Phi^T$ where $\Lambda$ is a diagonal matrix of the $m$ leading eigenvalues of $\Sigma$ and $\Phi$ the matrix of the corresponding eigenvectors. However, the computational complexity of generating the eigendecomposition scales the same as matrix inversion. \citet{Solin2020} propose an efficient method to approximate the eigenvalues and eigenvectors using Hilbert space methods, so we refer to it as a Hilbert Space Gaussian Process (HSGP). \citet{Riutort2023} provide further discussion of these methods.

Stationary covariance functions, including those in the Matern class like exponential and squared exponential, can be represented in terms of their spectral densities. For example, the spectral density function of the squared exponential function in $D$ dimensions is:
\begin{equation*}
S(\omega) = \sigma^2 (\sqrt{2\pi})^D \phi^D \text{exp}(-\phi^2 \omega^2/2)
\end{equation*}

Consider first a unidimensional space ($D=1$) with support on $[-c,c]$. The eigenvalues $\lambda_j$ (which are the diagonal elements of $\Lambda$) and eigenvectors $\phi_j$ (which form the columns of $\Phi$) of the Laplacian operator in this domain are:
\begin{equation*}
\lambda_j = \left( \frac{j\pi}{2L} \right)^2 
\end{equation*}
and
\begin{equation*}
\phi_j(s) = \sqrt{\frac{1}{c}} \text{sin}\left(  \sqrt{\lambda_j}(x+c) \right)
\end{equation*}
Then the approximation in one dimension is
\begin{equation*}
\mathcal{Z}(s) \approx \sum_{j=1}^m S\left(\sqrt{\lambda_j}\right)^{1/2}\phi_j(s)\beta_j
\end{equation*}
where $\beta_j \sim N(0,1)$. This result can be generalised to multiple dimensions. The total number of eigenvalues and eigenfunctions in multiple dimensions is the combination of all univariate eigenvalues and eigenfunctions over all dimensions. For example, if there were two dimensions and $m=2$ then there would be four multivariate eigenfunctions and eigenvalues equal to the four combinations over the two dimensions. The approximation is then as described above but with the multivariate equivalents.

The approximation can be used in the full model such that the intensity becomes
\begin{equation}
\label{eq:hsgpapprox}
    \lambda(S_i,t) = r(S_i,t)\exp\left(X(S_i,t)\gamma + (P^{\frac{1}{2}} \otimes \Phi\Lambda^{\frac{1}{2}}) \beta) \right)
\end{equation}
and where $\beta \sim N(0,I_{m^2})$. The matrix $\Phi$ does not depend on the covariance parameters are can be pre-computed, so only the product $\Phi\Lambda^{\frac{1}{2}}$ needs to be re-calculated during model fitting, which scales as $O(nm^2)$.

The performance of both the HSGP and the NNGP depends on $m$, either the number of basis functions or the number of nearest neighbours, respectively, and several other parameters. The other key value for the HSGP is the boundary condition $c$. \citet{Riutort2023} provide a detailed analysis of the HSGP for unidimensional linear Gaussian models, examining in particular how the choice of $c$ and $m$ affect performance of posterior inferences and model predictions. Here, we assume $c$ is relative to the maximum and minimum coordinate in each dimension (i.e. the space is scaled to $[-1,1]^2$). They find values of $m=15$ and $c=1.5$ to $2.5$ provide a good approximation for the models they consider with squared exponential covariance function, but also show that for higher values of $m$ we generally require larger values of $c$. For example, $m=20$ and $c=2$ provides similar performance to $m=30$ and $c=4$. 

\subsubsection{Gradient Approximation}
For maximum likelihood model fitting we can find the likelihood-maximising parameter values with fewer function evaluations using the gradient of the log-likelihood with respect to the model parameters. For the full log-likelihood and conditional on the random effects the gradients are given by:
\begin{align*}
    \frac{\partial \mathcal{L}}{\partial \beta} &= X^T(\mathbf{Y} - \boldsymbol{\lambda}) \\
    \frac{\partial \mathcal{L}}{\partial \theta} &= \frac{T}{2}\text{trace}\left(\Sigma_0^{-1} \frac{\partial \Sigma_0}{\partial \theta}\right) - \frac{1}{2}\sum_t \mathbf{z}_{[t]} \Sigma_0^{-1} \frac{\partial \Sigma_0}{\partial \theta}\Sigma_0^{-1} \Tilde{\mathbf{v}}_{[t]} \\
    \frac{\partial \mathcal{L}}{\partial \rho} &= \frac{N}{2}\text{trace}\left(P^{-1} \frac{\partial P}{\partial \rho}\right) - \frac{1}{2}\sum_t \mathbf{z}_{[t]} \Sigma_0^{-1}  \Tilde{\Tilde{\mathbf{v}}}_{[t]}
\end{align*}
where $\Tilde{\Tilde{\mathbf{v}}} = \mathbf{v}P^{-1}\frac{\partial P}{\partial \rho}P^{-1}$. For the NNGP we can approximate the gradients for the covariance parameters as:
\begin{align*}
    \frac{\partial \mathcal{L}}{\partial \theta} =& T\sum_{i=1}^N \frac{\partial D_{i,i}}{\partial \theta} D_{i,i}^{-2} + \frac{1}{2}\sum_t \mathbf{z}_{[t]} \frac{\partial A^T}{\partial \theta} D^{-1} (I - A) \Tilde{\mathbf{v}}_{[t]} \\
    & + \frac{1}{2}\sum_t \mathbf{z}_{[t]} (I-A)^T D^{-1} \frac{\partial A}{\partial \theta} \Tilde{\mathbf{v}}_{[t]} - \frac{1}{2}\sum_t \mathbf{z}_{[t]} (I-A)^T D^{-1}\frac{\partial D}{\partial \theta} D^{-1} (I-A) \Tilde{\mathbf{v}}_{[t]} \\
    \frac{\partial \mathcal{L}}{\partial \rho} &= \frac{N}{2}\text{trace}\left(P^{-1} \frac{\partial P}{\partial \rho}\right) - \frac{1}{2}\sum_t \mathbf{z}_{[t]} (I-A)^T D^{-1} (I-A)  \Tilde{\Tilde{\mathbf{v}}}_{[t]}
\end{align*}
The derivatives $\frac{\partial A}{\partial \theta}$ and $\frac{\partial D}{\partial \theta}$ can be calculated by differentiating the algorithms: $\Sigma_{0;\mathcal{N}_j,\mathcal{N}_j} \frac{\partial A_{j,\mathcal{N}_j}^T}{\partial \theta} = \frac{\partial \Sigma_{0;\mathcal{N}_j,j}}{\partial \theta} - \frac{\partial \Sigma_{0;\mathcal{N}_j,\mathcal{N}_j}}{\partial \theta}A_{j,\mathcal{N}_j}^T$ and $\frac{\partial D_{j,j}}{\partial \theta} = \frac{\partial\Sigma_{0;j,j}}{\partial \theta} -  \frac{\partial A_{j,\mathcal{N}_j}}{\partial \theta}\Sigma_{0;\mathcal{N}_j,j} - A_{j,\mathcal{N}_j} \frac{\partial \Sigma_{0;\mathcal{N}_j,j}}{\partial \theta}$.

For the HSGP, we differentiate the log-likelihood with the intensity given by Equation (\ref{eq:hsgpapprox}):
\begin{align*}
     \frac{\partial \mathcal{L}}{\partial \theta} &= \text{trace}\left( (\mathbf{Y} - \boldsymbol{\lambda})^T \beta^T (P^{\frac{1}{2}} \otimes \Phi \frac{\partial \Lambda^{\frac{1}{2}}}{\partial \theta}) \right)
\end{align*}

These gradient approximations are more complex to calculate than the Gaussian Process approximations themselves. The quality of the approximations may also hinder convergence. So we provide both derivative-free and quasi-Newton optimisers for maximum likelihood estimation using the approximations.

\subsection{Model for Spatially-Aggregated Case Data}
The models discussed so far have assumed that the precise location of each case is known. However, in many cases we will have instead case count aggregated to census or administrative areas for reasons such as preserving anonymity. These areas are almost always irregularly shaped and have a large variability in area. As a result, we cannot necessarily assume that the underlying latent process is approximately constant within each area, as we do with the regular lattice, which can cause bias. We can instead extend the LGCP model described above to account for the irregular areas onto which case counts are aggregated.  

Similar to \citet{Li2012}, we notate the count data for this `region' model as $Y(R_j,t)$ where $R_j$ is the $j$th region (i.e. irregular area) of $r$ total regions. We may have covariates for both the region, $X_r(R,t)$ and cell $X_s(S,t)$. The intersection areas between the regions and the computational grid are $Q_{ij} = R_j \cap S_i$ where there are $q$ non-empty intersection areas. The intensity for area $Q_{ij}$ is:
\begin{equation}
\label{eq:intersectionintens}
    \lambda(Q_{ij},t) = w_{ij}r(R_j,t)\exp\left(X_r(R_j,t)\gamma_r + X_s(S_i,t)\gamma_s + Z(S_i,t)\right)
\end{equation}
where the weights are $w_{ij} = \frac{\vert \vert Q_{ij} \vert \vert}{\vert \vert R_j \vert \vert}$. The region level counts are $Y(R_j,t) = \sum_{i\in\mathcal{O}_j} Y(Q_{ij},t)$ where $\mathcal{O}_j$ are the indexes of the grid cells that overlap with $R_j$. The counts in the intersections are not observed but given our assumptions, we can write the intensity for $R_j$ at time $t$ as the sum of the intensities of the intersections in the region:
\begin{align}
    \begin{split}
    \label{eq:regionintens}
        \lambda(R_j,t) = r(R_j,t)\exp\left(X(R_j,t)\gamma) \right)\left[ \sum_{i \in \mathcal{O}_j} w_{ij} \exp\left(X_s(S_i,t)\gamma_s +  Z(S_i,t) \right) \right]
    \end{split}
\end{align}
and the latent field is then specified as above. 

\citet{Li2012}, building on work in \citet{Li2012b}, describe a data augmentation scheme for estimating the LGCP with aggregated case data. Their approach models the counts at the intersection level. As the locations of each case within a region are not known, and so it is not known which computational grid cell to add them to, they assign a case to a grid cell within the region with probability proportional to the $w_{ij} \exp\left( Z(S_i,t) \right)$ on each iteration of the MCMC sampling and then model the intersection intensities. Despite proposing a INLA approach to modelling the random field, the method is computationally intensive given the number of intersections and the need for a data simulation step. In \pkg{rts2} we instead aim to model the counts at the region level directly and instead aggregate the intensities of the intersections, which is a more efficient method of sampling from or fitting an equivalent model. We assume that the population density and any covariates for the regions are constant throughout the area, then the model is then amenable to the fitting strategies discussed in this article. 

We also define three different design matrices that are used in later calculations relating to this ``region model''. First, we let $A$ be the ``region to intersection'' design matrix with dimension $q \times r$. In each row $q$ there is a one in the column corresponding to the region of the intersection. Similarly, $B$ is a $q \times n$ matrix that provides a mapping from the computational grid to the intersections. Both matrices are sparse. Finally, the ``grid to region design matrix'' maps from the grid to the regions and is equal to $C = B^TA$. If $\boldsymbol{\mu}_S$ is $nt$ length vector with elements equal to $\exp\left(X_s(S_i,t)\gamma_s \right)$, $\boldsymbol{\mu}_Z$ is $nt$ length vector with elements equal to $\exp\left(Z(S_i,t) \right)$ and $\boldsymbol{\mu}_R$ a $rt$ length vector with elements $r(R_j,t)\exp\left(X(R_j,t)\gamma) \right)$ such that the regional intensities can be written as $\boldsymbol{\mu}_R \odot (C (\boldsymbol{\mu}_S \odot \boldsymbol{\mu}_Z))$ and $\odot$ represents the Hademard product.

\subsection{Bayesian model fitting}
There are several approaches to estimating the LGCP in a Bayesian framework. Letting $\Theta = [\beta, \theta, \sigma, Z]$ represent all the model parameters, the Bayesian approach aims to draw samples from the posterior 
\begin{equation*}
f_{\Theta|\mathcal{D}}(\Theta|\mathcal{D}) \propto f_{\mathcal{D}|\Theta}(\mathcal{D}|\Theta) f_\Theta(\Theta)
\end{equation*}
Markov Chain Monte Carlo (MCMC) methods are often used. In the context of the LGCP, there has been some discussion of the choice of the proposal density for the Metropolis-Hastings algorithm. \citet{Moller1998} propose a Langevin kernal for, which is known as the Metropolis-adjusted Langevin algorithm (MALA). MALA is a special case of Hamiltonian Monte Carlo (HMC), which uses Hamiltonian dynamics to construct proposals for the Metropolis-Hastings algorithm \citep{Carpenter2017}. MALA was used by \citep{Taylor2013} for fitting an LGCP. HMC is potentially more efficient than MALA though \citep{Teng2017}.  

MCMC algorithms have the advantage of providing consistent estimation of various characteristics of the posterior distribution, including well calibrated probabilities. However, it is computationally intensive, and so may provide a further limit on the granularity of predictions in settings where time is a constraint. Several deterministic methods for approximate Bayesian inference have been proposed to try to overcome this limitation. One such method is Variational Bayes (VB) using an approximating density $r(\Theta; \phi)$ parameterised by $\phi$. VB minimises the Kullback-Leibler (KL) divergence from the approximation density to the posterior:
\begin{equation*}
    \min_\phi \text{KL}(r(\Theta; \phi) \vert \vert f_{\Theta| \mathcal{D}}(\Theta| \mathcal{D}))
\end{equation*}
For the LGCP, the KL divergence lacks an analytical form, so a proxy is used, the evidence lower bound:
\begin{equation*}
    \mathcal{L}(\phi) = E\left[ \log f_{\Theta,\mathcal{D}}(\Theta,\mathcal{D}) \right] - E\left[ \log r(\Theta; \phi) \right]
\end{equation*}
where the expectations are with respect to the prior densities. The minimisation problem is then to find $\phi$ that maximises $\mathcal{L}(\phi)$ such that the approximation density is in the support of the posterior. One of the difficulties with VB is deriving an appropriate family of approximating densities and then determining the appropriate model-specific quantities for the VB problem. \citet{Teng2017} give a specific derivation for the full LGCP. However, an alternative approach is to use automatic differentiation methods that provides a general algorithm for probabilistic models, which enables use of the Gaussian process approximations discussed here. 

\subsection{Maximum Likelihood Estimation}
The Markov Chain Monte Carlo Maximum Likelihood (MCMCML) and Stochastic Approximation Expectation Maximisation (SAEM) algorithms can fit the LGCP and both are implemented in the \pkg{rts2} package. The negative log-likelihood of (\ref{eq:lgcplik}) can be rewritten as:
\begin{equation}
\label{eq:loglik}
    -\log L(\gamma,\theta | \mathcal{D})  = \mathcal{L}(\gamma,\theta | \mathcal{D}) = E_z\left[ 
 -\log f_{Y|z}(Y(S_i,t)|\gamma,\mathbf{z}) \right] + E_z\left[-\log f_z (\mathbf{z} | \theta) \right]
\end{equation}
both components of this log-likelihood can be estimated using MCMC and then minimised.

\subsubsection{Markov Chain Monte Carlo Maximum Likelihood}
MCMCML algorithms for mixed models are described in detail by \citet{McCulloch1997} and implemented for GLMMs in general in the \pkg{glmmrBase} package for R, which we build on for model fitting in this package. The MCMCML algorithms have three steps that are implemented on each iteration. On the $k$th iteration:
\begin{enumerate}
    \item Sample $m_k$ latent effects $\mathbf{z}^{(k)}$ from the distribution $\mathbf{z}|Y$ using MCMC.
    \item Obtain new estimates $\hat{\gamma}^{(k)}$ from the likelihood of $Y$ conditional on $\mathbf{z}^{(k)}$ by minimising $ \mathcal{L}(\gamma) = E_z\left[ 
 -\log f_{Y|z}(Y(S_i,t)|\gamma,\mathbf{z}) \right] \approx \frac{1}{m_k} \sum_{i = 1}^{m_k} -\log f_{Y|z}(Y(S_i,t)|\gamma,\mathbf{z}^{(k,i)})$. 
    \item Obtain new estimates of the covariance parameters $\hat{\theta}^{(k)}$ by minimising $\mathcal{L}(\theta) = E_z\left[-\log f_z (\mathbf{z} | \theta) \right] \approx \frac{1}{m_k}\sum_{i=1}^{m_k} -\log f_z (\mathbf{z}^{(k,i)} | \theta) $.
\end{enumerate}
The steps are then repeated until convergence, see below. 

\subsubsection{Stochastic Approximation Expectation Maximisation}
One way to improve the efficiency of the MCMCML algorithm is to make use of all MCMC samples at each iteration and only draw a small number $m_k$ each time. SAEM is a Robbins-Munroe approach for minimising functions of the form (\ref{eq:loglik}) \citep{Jank2006}. In particular, the log-likelihood for $\gamma$ is approximated on each iteration as:
\begin{equation*}
    \hat{\mathcal{L}}^{(k+1)}(\gamma) = \hat{\mathcal{L}}^{(k)}(\gamma) + \alpha_k \left( 
 \frac{1}{m_k} \sum_{i = 1}^{m_k} -\log f_{Y|z}(Y(S_i,t)|\gamma,\mathbf{z}^{(k,i)}) - \hat{\mathcal{L}}^{(k)}(\gamma) \right)
\end{equation*}
and similarly for $\theta$, and where $\alpha_k \propto \left( \frac{1}{k} \right)^\alpha$ for $\alpha \in [0.5,1)$.  The log-likelihood in this algorithm is a convex combination of all the information from prior iterations. This algorithm converges with constant value of $m_t$ REF. The only choice required of the user is of the values of $\gamma_t$. Small values of $\alpha$ lead to faster convergence but higher Monte Carlo error as the past is more quickly ``forgotten'', while larger values of $\alpha$ lead to slower convergence but lower Monte Carlo error \citep{Jank2006}. Convergence can be improved using Polyak-Ruppert averaging \citep{Polyak1992}, where the estimate for the log-likelihood for $\gamma$ is:
\begin{equation*}
    \Tilde{\mathcal{L}}^{(k)}(\gamma) = \frac{1}{k}\sum_{i = 1}^k \Tilde{\mathcal{L}}^{(k)}(\gamma)
\end{equation*}

\subsubsection{Stopping criteria}
There are two approaches for stopping the stochastic algorithms. The first considers the differences, or maximum absolute difference, between point estimates of the parameters on successive iterations, and stops the algorithm when the difference is below some threshold. A running mean difference can be used to avoid premature termination. A second method is to compare the log-likelihood values between iterations since:
\begin{equation*}
    \Delta \mathcal{L}(k | k - 1)  = \mathcal{L}^{(k)}(\hat{\gamma}^{(k)},\hat{\theta}^{(k)}) - \mathcal{L}^{(k)}(\hat{\gamma}^{(k-1)},\hat{\theta}^{(k-1)}) < 0
\end{equation*}
when the algorithm has converged, where $\mathcal{L}^{(k)}(\hat{\gamma}^{(k)},\hat{\theta}^{(k)}) = \mathcal{L}^{(k)}(\hat{\gamma}^{(k)}) + \mathcal{L}^{(k)}(\hat{\theta}^{(k)})$. While an expectation maximisation algorithm is guaranteed to improve the log-likelihood, the Monte Carlo error in a stochastic approach can lead to reductions in the log-likelihood even at convergence. \citet{Caffo2005} proposed considering an upper bound on this difference:
\begin{equation*}
    U(k | k - 1) = \Delta \mathcal{L}(k | k - 1) + z_\delta \hat{\sigma}_\Delta < 0
\end{equation*}
where $z_\delta$ is a standard normal quantile at $\delta$ and $\hat{\sigma}^2_\Delta$ is an estimate of the variance of $\Delta \mathcal{L}(k | k - 1)$. When satisfied, there would be a probability of $\delta$ of convergence.

The number of steps on each iteration $m_k$ also does not need to be constant. The first iterations are needed for the estimates to converge, with later iterations needed to refine those estimates. \citet{Caffo2005} propose modifying the number of samples as:
\begin{equation*}
    m_{k+1} = \max \left( m_k, \mathcal{L}^{(k)}(\gamma,\theta) \frac{(z_\delta + z_\epsilon)^2}{\Delta \mathcal{L}(k | k - 1)^2}  \right)
\end{equation*}
where $\epsilon$ is the type 2 error rate. We implement MCMCML with both adaptive and non-adaptive step sizes.
 
\subsubsection{Computational efficiency}
The MCMCML approach may provide some speed advantages over full MCMC Bayesian methods in some scenarios. The first step can be modified: rather than sampling $\mathbf{z}$ we can instead sample $\mathbf{z}^* = (R \otimes L)^{-1}\mathbf{z}$ where $L$ is the Cholesky decomposition of $\Sigma_0 = LL^T$ and $R$ is the decomposition of $P$. In particular, we use MCMC to sample $\mathbf{z}^*$ from:
\begin{align}
    \begin{split}
    \label{eq:cholmodel}
        \lambda(s_i,t) &= r(S_i,t)\exp (X(s_i,t)\hat{\beta}^{(k)} + (R \otimes L)\mathbf{z}^{*(k)}) \\
        \mathbf{z}^{*(k)} &\sim N(0,I)
    \end{split}
\end{align}
The second step is equivalent to fitting a generalised linear model and scales with $O(nTQ)$. The third step is the most computationally intensive, as discussed above, and we can use one of the approximations. For the NNGP, model (\ref{eq:cholmodel}) we can generate an approximation to the Cholesky decomposition as $L \approx (I-A)^{-1}D^{1/2}$. The lower triangular matrix $(I-A)^{-1}$ can be obtained using a forward substitution algorithm shown in Algorithm \ref{alg:approxchol}, the outer loop of which can be parallelised. Thus, generation of the approximate Cholesky factor scales as $O(nm^2)$ after the cost of calculating $A$ is bourne, which is $O(nm^3)$.

\begin{algorithm}
\caption{Calculation of $(I-A)^{-1}D^{1/2}$}
\label{alg:approxchol}
\begin{algorithmic}
 \State Let $\Tilde{L}$ be an $n \times n$ lower triangular matrix
 \ForAll{$i \in 1,...,n$}
 \ForAll{$k \in i,...,n$}
 \State Let $l = A_{\mathcal{N}_i,i}^T \Tilde{L}_{\mathcal{N}_j,k}$
  \If{$i=k$}
    \State $\Tilde{L}_{i,k} = (1 + l)\sqrt{D_{kk}}$
 \Else
    \State $\Tilde{L}_{i,k} = l\sqrt{D_{kk}}$
 \EndIf
 \EndFor
 \EndFor
\end{algorithmic}
 \end{algorithm}

For the HSGP, we instead use $L' = \Phi\Lambda^{1/2}$ as a square root of $\Sigma_0$ in place of $L$. The calculations for $\Phi$ and $\Lambda$ are given above. 

\subsubsection{Region model}
To modify Equation (\ref{eq:cholmodel}) for the region model, we return to the expression for the regional intensity as $\boldsymbol{\mu}_R \odot (C (\boldsymbol{\mu}_S \odot \boldsymbol{\mu}_Z))$. We can rewrite this as $C^*\boldsymbol{\mu}_Z$ where $C^*$ is equal to the weight matrix $C$ where the $j$th column is multiplied by the $j$ th element of $\boldsymbol{\mu}_S$ and where the $i$th row is multplied by the $i$th element of $\boldsymbol{\mu}_{R}$. For the MCMC sampling step, the matrix $C^*$ can be pre-calculated and the random effects sampled from 
\begin{align}
    \begin{split}
    \label{eq:cholmodel}
        \lambda(r_i,t) &= C^*\exp (L\mathbf{z}^{*(k)}) \\
        \mathbf{z}^{*(k)} &\sim N(0,I)
    \end{split}
\end{align}

\subsubsection{Standard errors}
There are two approaches we can use to estimate the standard errors for the model parameters $\gamma$ (or $\gamma_r$ and $\gamma_s$). First, if $\Omega$ is the covariance matrix for $Y$, then we can use the estimated information matrix $M = X^T\Omega_{\hat{\theta}}^{-1}X$. The standard errors for $\gamma$ are then given by the square root of the diagonal of $M^{-1}$. Estimating $\Omega$ we can use a first-order approximation based on the marginal quasi-likelihood \citep{breslow1993approximate}:
\begin{equation*}
    \Omega = W(\hat{\gamma}) + \Sigma(\hat{\theta})
\end{equation*}
where $W(\hat{\gamma})$ is a diagonal matrix with elements $1/\lambda(S,t)$, which are equivalent to the inverse GLM iterated weights for the Poisson-log model and is evaluated at the estimated parameter values. The matrix $\Sigma(\hat{\theta})$ is also evaluated at the estimated values of the covariance parameters. 

For regional data models with intensity given by Equation (\ref{eq:regionintens}), the intensity is now a more complex sum of exponentials and so we cannot use the Poisson-log model information matrix directly. However, we can instead use the intersection-level model, which is log-linear in all its terms, and use the same information matrix approach with the estimated parameters from the region model to obtain the standard errors. This could prove very costly computationally however, as the number of intersection areas may be very large relative to the size of the computational grid and number of regions, and we require inversion of the covariance matrix of size $QT$. The auto-regressive covariance structure provides a way of reducing the cost of obtaining this inverse. In particular, the intersection linear predictors are given by:
\begin{equation*}
    \boldsymbol{\mu}_Q =  A X_r\gamma_r + B X_s\gamma_s + A_{s \rightarrow q}Z + log(\mathbf{w}) 
\end{equation*}
and the weight matrix $W'(\hat{\gamma})$ is a diagonal matrix with entries $1 / \exp(\boldsymbol{\mu}_Q)$. We let $W'_{t}(\hat{\gamma})$ be the sub-matrix of $W'(\hat{\gamma})$ corresponding to time period $t$, then the grid cell intensities for time period $t$ are given by $\boldsymbol{\lambda}_{G,t} = \exp(B^TW'_t(\hat{\gamma}))$
\begin{equation*}
    \Omega' = (P^{-1} \otimes \Sigma_0^{-1}) + \text{diag}(\boldsymbol{\lambda}_{G,t})
\end{equation*}
Using the notation $\Sigma_{0,st}$ to refer to the rows of $\Sigma_0$ in time period $s$ and the columns in $t$, we can then obtain the inverse covariance matrix at the intersection level as:
\begin{equation*}
    \Omega^{'-1} = \begin{bmatrix}
        W_{1}^{'-1} & & \\
          & W_{2}^{'-1} & \\
          & & \ddots 
    \end{bmatrix} - 
    \begin{bmatrix}
        W_{1}^{'-1}B \Sigma_{0,11}^{-1}B^TW_{1}^{'-1} & W_{1}^{'-1}B\Sigma_{0,12}^{-1}B^TW_{2}^{'-1}  & \hdots\\
          W_{2}^{'-1}B\Sigma_{0,21}^{-1}B^TW_{1}^{'-1} & W_{1}^{'-1}B\Sigma_{0,22}^{-1}B^TW_{2}^{'-1} & \hdots \\
         \vdots &\vdots & \ddots 
    \end{bmatrix}
\end{equation*}
Then the information matrix is given as:
\begin{equation*}
    M = \begin{bmatrix}
        AX_r \\
        BX_s
    \end{bmatrix}^T
    \Omega_Q^{-1} \begin{bmatrix}
        AX_r \\
        BX_s
    \end{bmatrix}
\end{equation*}
As the matrix $W'$ is diagonal and the design matrices are sparse, much of the computation is relatively cheap. The most expensive operation is the inversion of $\Sigma_0$, which is much smaller in dimension than inverting $\Omega'$ directly.

The approximation above may fail when using the HSGP approximation, as $\Sigma$ and hence $\Omega$ will be rank deficient and hence not invertible. Using $\Phi \Lambda^{-\frac{1}{2}} \Phi^T$ also does not provide a good approximation to the inverse. Future versions will aim to provide a good approximation to standard errors of fixed effect parameters for region data models, with HSGP, and MCMCML.

\section{Related Software and Methods}
\label{sec:software}
Software packages for fitting an LGCP are relatively few. In the R ecosystem, we are aware of two dedicated packages. First, \pkg{lgcp}\citep{Taylor2013,Taylor2015} provides maximum likelihood and Bayesian MCMC estimation of the LGCP model. However, it uses only the full model and while there are some features to improve computational time, it suffers from long running times for all but the smallest applications. The temporal structure of the implemented model does not scale linearly with time either. Furthermore, it makes use of a no-longer supported package and standard for spatial data (\pkg{sp}) and does not make use of state-of-the-art Bayesian methods. We would also argue that the user interface may present a barrier to use in a surveillance system as it is relatively complex. The \proglang{R} package \pkg{inlabru} \citep{Bachl2019} provides a set of tools to approximate Bayesian posteriors the LGCP using integrated nested Laplacian approximation (INLA), a popular and highly effcient Bayesian approximation. INLA (made available in R more generally through \pkg{R-INLA} \citep{Lindgren2015}) approximates the posterior means of model parameters, and in this case spatially-continuous Gaussian processes, using Gaussian Markov random fields (GMRFs). A comprehensive overview is provided in \citet{Rue2005} Chapter 5. We do not include INLA given the availability of \pkg{inlabru}. The quality of INLA in terms of quantifying uncertainty beyond posterior means may also be less than needed for surveillance applications. \citet{Taylor2014} compare MCMC (specifically MALA) and INLA using the methods in \pkg{lgcp} and \pkg{inlabru} and conclude that while MCMC does take much longer, it offers much better predictive accuracy. However, the scale of the problems examined may be considered relatively small, for larger grids and multiple time periods, MCMC may be prohibitively slow. \citet{Teng2017} come to similar conclusions in a broader comparison of MCMC and other methods. We are not aware of any packages specifically designed for LGCPs, or real-time surveillance more generally, for other software environments. We are also not aware of any other implementation of what we have termed the `region model' or of NNGP or HSGP with LGCPs.

One may also fit LGCPs using more general mixed model software. For example, provided one generate the appropriate grid data, the R package \pkg{glmmTMB} offers Laplace approximation model fitting for GLMMs including Poisson models with spatial covariance structure. One can extract the processed grid data generated by \pkg{rts2} for use in other packages. For full maximum likelihood model fitting one can use \pkg{glmmrBase}. For Bayesian MCMC the Stan probabilistic programming language can fit any of the models described here. The \pkg{rts2} package uses \pkg{glmmrBase} and \proglang{Stan} for model fitting. In Stata and SAS, one could theoretically also use the GLMM fitting functionality to fit these models, however, as with the other methods, they would be impractically slow for even moderately sized grids. We are not aware of any specific implementation of LGCP models in other software environments.

\section{rts2 package}
We now describe the functioning of the package in R and provide some comparisons between the different methods. MCMC sampling uses Stan. There are two main packages providing an interface with Stan available for \proglang{R}: \pkg{rstan} \citep{Carpenter2017} and \pkg{cmdstanr}. The latter package provides a lightweight interface to Stan as an external program and so is up to date with the latest version of Stan (2.36). 

The main functionality of \pkg{rts2} is provided by the \code{grid} class, which uses the \pkg{R6} package's object-orientated class system. An object-orientated method is preferred here, as much of the functionality of the package involves repeatedly manipulating the same data object. As we demonstrate below, use of object-orientated class system simplifies the syntax relative to the more standard \proglang{R} functional methods in these cases. Geographic data uses the \pkg{sf} package. Initialisation of a new \code{grid} object can use either a single polygon describing the boundary of an area of interest, or a set of polygons in the case of the region data model described above.

\subsection{Point data}
We use simulated data for this first example, which are provided in the package, as real-world location identifying data are typically not anonymised. These data, along with the relevant spatial data objects, are provided in the replication materials. The boundary is also provided and represents the city of Birmingham, UK along with a set of covariate data at the middle layer super output area (MSOA) level, which is the second smallest census tract area. 

\begin{figure}
    \centering
    \includegraphics[width=\textwidth]{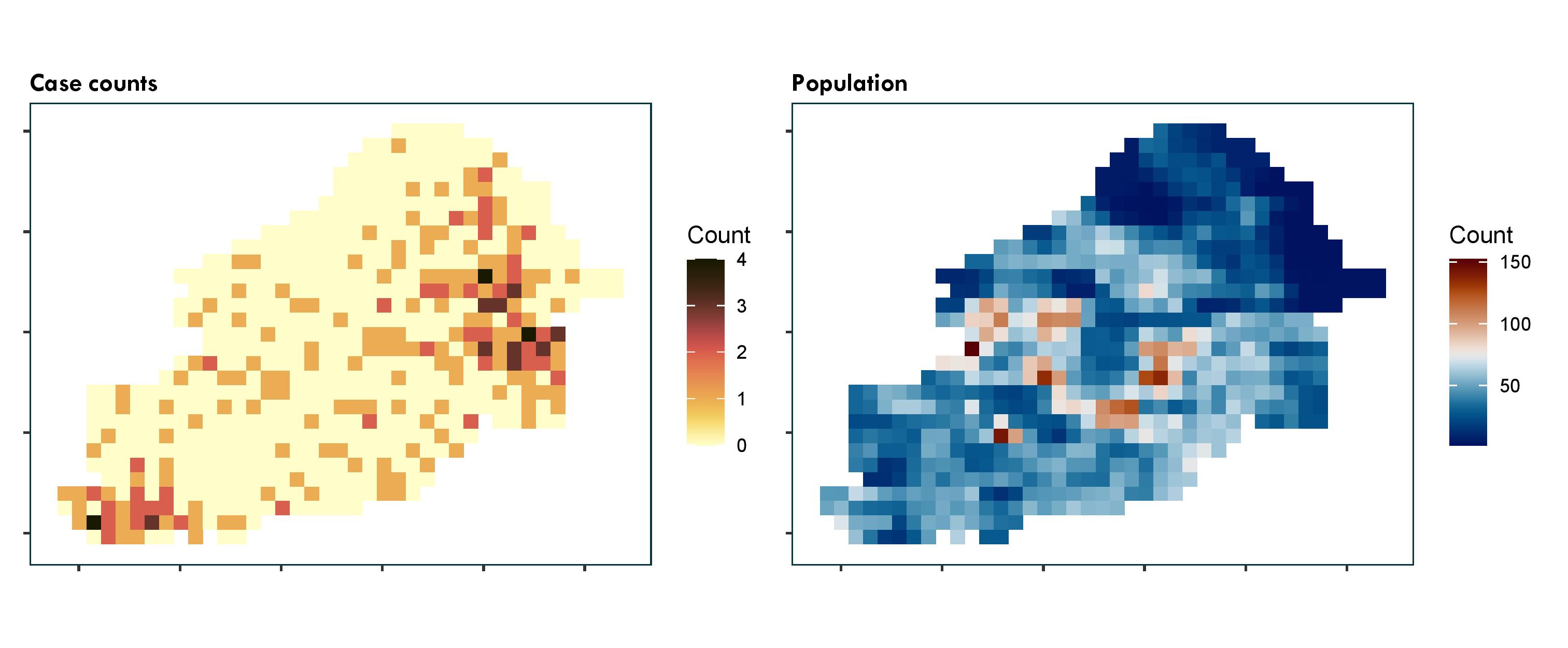}
    \caption{Example simulated data for the city of Birmingham, UK. Left panel: case counts of point data mapped to the computational grid. Right panel: population density mapped to the computational grid.}
    \label{fig:single-example}
\end{figure}

To instantiate a new \code{grid} object we require the boundary and the cell size of the grid:
\begin{CodeChunk}
\begin{CodeInput}
g1 <- grid$new(boundary,cellsize=0.008)
\end{CodeInput}
\end{CodeChunk}
The next step is to map the point data to the grid to create the counts. We provide the convenience function \code{create\_points()} to generate \pkg{sf} point data from an ordinary R data frame with location and possibly time data. Then, the member function \code{points\_to\_grid()} will aggregate the case counts, for example:
\begin{CodeChunk}
\begin{CodeInput}
g1$points_to_grid(point_data = create_points(y,
                                             pos_vars = c("Y","X"),
                                             t_var = c("t")),
                  t_win = "month",
                  laglength = 1)
\end{CodeInput}
\end{CodeChunk}
For spatio-temporal analysis, we could determine here what size time step we want (day, week, or month) and how many previous periods to include (\code{laglength}). The choice of lag length is similar to the choice of cell size, we want it to be long enough to provide the best predictions of current incidence, but as short as possible otherwise to minimise the required computational resources. Note that the computational time is linear in the number of time periods. The function adds columns to the grid object (labelled \texttt{t*} for spatio-temporal or \texttt{y} for spatial-only). Figure \ref{fig:single-example} top-left panel shows the aggregated case counts, which can be plotted using \code{g1$plot("y")}.

\subsection{Overlaying covariate data}
We can include covariates in our model. The aim of including covariates in surveillance applications is not typically to provide inference about the effect of any particular covariate but to improve predictions or to generate conditional, e.g. age-adjusted, predictions of relative risk. For example, by including day of the week as a covariate we can capture differences in hospital attendances and admissions resulting from individual behaviour not related to variations in disease incidence. At a minimum we would suggest including population density to enable population standardised predictions. Covariates can be added to the grid data using the member function \texttt{add\_covariates()}. There are three main types of covariate.

\subsubsection{Spatially-varying only covariates}
Where the time step is short, such as days, it is unlikely there will be data that varies by time step over the area of interest. For example, population density data will be static over the area of interest. To add spatially-varying covariates to the grid data we require an \texttt{sf} object with polygons encoding the covariate data. The \texttt{add\_covariates()} function then, for each grid cell, takes an average of the values in the overlapping polygons, weighted by either the size of the area of overlap or the population, if population density data are available. A good source of population density predictions is WorldPop (at www.worldpop.org), which will provide rasters at 100m or 1km resolution for all countries.

First, we add the population density:
\begin{CodeChunk}
\begin{CodeInput}
g1$add_covariates(msoa,
                  zcols="popdenshect",
                  weight_type="area")
\end{CodeInput}
\end{CodeChunk}
where \texttt{msoa} is the MSOA data as an \code{sf} object. Figure \ref{fig:single-example} top right panel shows the population density.

\subsubsection{Temporally-varying only covariates}
The example includes only a single-period, but in other cases with multiple time periods we may have both temporally and spatio-temporally varying covariates. Some covariates, like day of the week, vary over time but are the same for all grid cells in each time period. To add a temporally-varying covariate we need to produce a data frame with a column for time period (from 1 to \code{laglength}) and then columns with the values of the covariates. To obtain day of the week we can use the member function \code{get\_dow()}, which will return a data frame mapping the date of each time period to a day of the week. For example,
\begin{CodeChunk}
\begin{CodeInput}
df_dow <- grid_data$get_dow()
print(df_dow)
\end{CodeInput}
\end{CodeChunk}
will return the data frame and then these columns can then be added to the main grid data:
\begin{CodeChunk}
\begin{CodeInput}
grid_data$add_covariates(df_dow,
                        zcols=c("dayMon","dayTue",
                                "dayWed","dayThu","dayFri",
                                "daySat","daySun"))
\end{CodeInput}
\end{CodeChunk}

The member function \code{add\_time\_indicators()} will generate fixed effect indicator variables for each time period in multi-period data. These will be labelled \code{time1i}, \code{time2i}, and so forth.

\subsubsection{Spatially- and temporally-varying covariates}
In some cases there may be covariates that vary over time and space. There are two ways of adding these into the main grid data. Both ways need to generate, for each covariate, as many columns as there are time periods. The column names will be the covariate name followed by the time period index.

If the covariate data are included in different data frames for each time period, then we can repeat the process above for each time period and add a label using the \texttt{t\_label} argument:
\begin{CodeChunk}
\begin{CodeInput}
grid_data$add_covariates(msoa_t1,
                         zcols=c("covA","covB",...),
                         t_label = 1)
\end{CodeInput}
\end{CodeChunk}
Alternatively, if the value of the covariates are all included in different columns of the same data frame then they can be added at the same time, ensuring we keep to the same naming convention of covariate name and time index as a string:
\begin{CodeChunk}
\begin{CodeInput}
grid_data$add_covariates(msoa,
                         zcols=c("covA1","covA2","covA3",...))
\end{CodeInput}
\end{CodeChunk}

\subsection{Model fitting}
The two model fitting functions are \code{lgcp\_bayes()} and \code{lgcp\_ml()} for Bayesian and maximum likelihood methods, respectively. The arguments to the two functions are highly similar. The functions described above add the case counts and covariates to the grid data. The model fitting functions require specification of which covariates to include, the model to use, and any approximation and fitting parameters. For example, to use the exponential covariance function with the NNGP approximation with ten nearest neighbours, we can use the following code. We first re-order the computational grid using the `minimax' ordering.
\begin{CodeChunk}
\begin{CodeInput}
g1$reorder("minimax")
g1$lgcp_ml(popdens = "popdenshect",
           approx = "nngp",
           m= 15)
\end{CodeInput}
\end{CodeChunk}
The model fit can be returned from this call, but it is also stored internally and can be returned using \code{g1$model\_fit()}, which will return an S3 object of class \code{rtsFit}. 

\begin{figure}
    \centering
    \includegraphics[width=\textwidth]{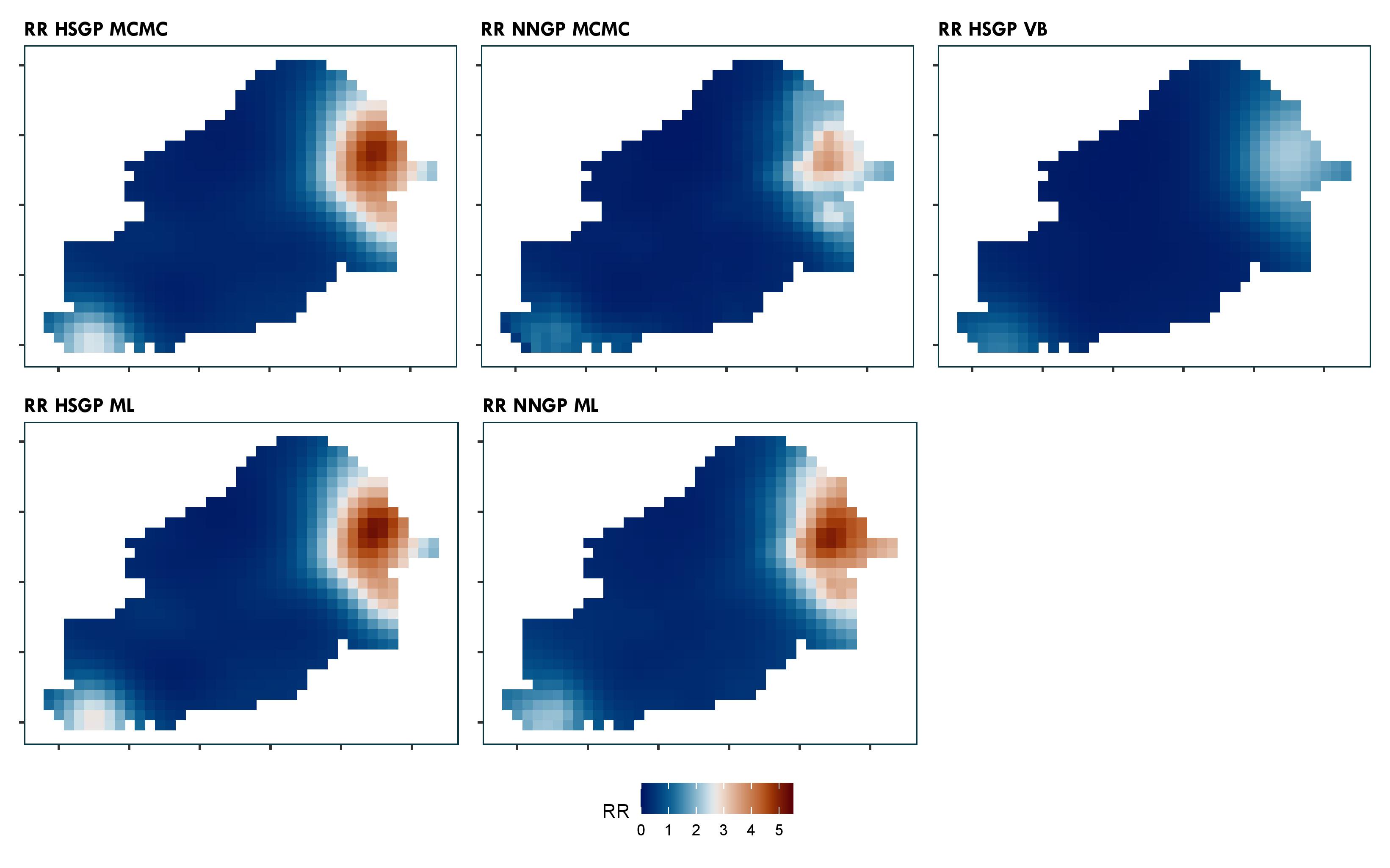}
    \caption{Example model fits using five methods: MCMC, Variational Bayes (VB) and MCMC Maximum likelihood, along with HSGP and NNGP approximations. The HSGP model fits used 15 basis function per dimension and the NNGP used 25 nearest neighbours. }
    \label{fig:single_example}
\end{figure}

Figure \ref{fig:single_example} shows the predicted relative risk from five different model fitting methods. We discuss extraction and plotting of model fits in the next section. Code to generate these model fits is provided in the replication materials. We note that we exclude a method using variational Bayes and NNGP, as this combination of approximations does not appear to produce reliable predictions. The predictions are qualitatively similar, although there are notable differences in the smoothness and magnitude of the relative risk surface, which result from different choices about approximation parameters and other aspects of the analysis.

\subsection{Region data}
To illustrate the use of the regional data model we use publicly available crime data from the UK. Monthly counts of different types of crime are available at the middle-layer super output area level from crime.uk. We extract the monthly counts for 2022 for the city of Birmingham, UK for burglary. These data are available in the package.

If a new \code{grid} class object is instantiated with data comprised on multiple polygons, it assumes a regional data model. The provided polygons are assumed to represent all the areas of interest. Generally, these should be a division of a contiguous area, although this is not a strict requirement. Since the context for a regional data model is that the data are only available in their aggregated format, the data used to initialise the new \code{grid} object should already contain these counts. Column names should be in the same format as generated for point data described above: if there is only a single time period then the counts should be in a column named \code{y}, otherwise the columns should be labelled \code{t1}, \code{t2}, \code{t3}, ... and so forth in chronological order. As with the point data we must also provide a cell size for the computational grid. Covariates must also be included in the regional data object with the same naming conventions for multi-period models. One can add grid level covariates from another source using the \code{add_covariates()} function as described above. 

\begin{CodeChunk}
\begin{CodeInput}
g2 <- grid$new(msoa,cellsize = 0.008) # create new region grid 
g2$reorder("minimax") # re-order for NNGP
g2$add_time_indicators()
\end{CodeInput}
\end{CodeChunk}

We do not include any additional covariates in this example beyond the time indicators. Model fitting uses the same commands as for the basic spatial data model described above. We fit the model with maximum likelihood and both NNGP and HSGP approximations:

\begin{CodeChunk}
\begin{CodeInput}
g2$lgcp_ml("pop", # population density
            covs = paste0("time",2:12,"i"), # time period indicators
            approx = "nngp", # approximation
            m = 15, # number of nearest neighbours
            iter_warmup = 100, # iterations for the MCMC step
            iter_sampling = 50) 

g2$lgcp_ml("pop",
           covs = paste0("time",2:12,"i"),
            approx = "hsgp", 
            m = 10, 
            iter_warmup = 150,
            iter_sampling = 50)
\end{CodeInput}
\end{CodeChunk}
A relative small number of sampling iterations are specified, as the default fitting procedure is SAEM, which will incorporate all prior sample information in the estimation in each step.

\subsection{Analysing the output}
\begin{figure}
    \centering
    \includegraphics{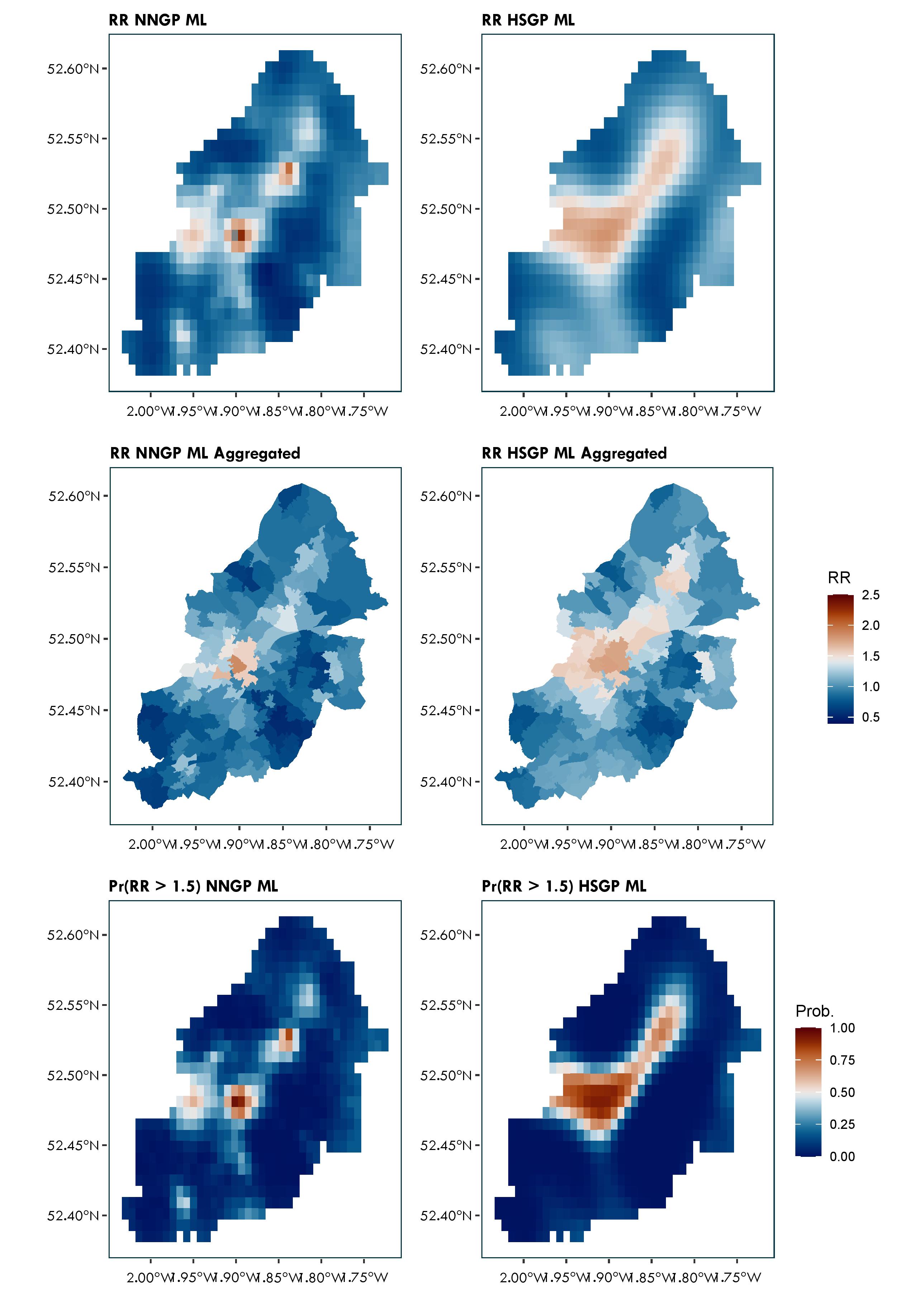}
    \caption{Model fitting outputs for the Birmingham crime example. The left column provides outputs from a NNGP maximum likelihood model fit with 15 nearest neighbours; the right column is HSGP maximum likelihood with 10 basis functions per dimension. The top row is the predicted relative risk, the middle row the aggregated relative risk, and the bottom row the probability that the relatve risk is greater than 1.5.}
    \label{fig:bham_example}
\end{figure}

Figure \ref{fig:bham_example} shows several outputs from the models fitted in the previous section. Here, we describe how to extract these outputs.

\subsubsection{Predictions}
The output from the model enables us to generate four main predictions across our area of interest:
\begin{enumerate}
    \item Predicted total incidence.
    \item Predicted incidence per head of population.
    \item The relative risk, i.e. the relative difference between predicted and expected incidence. For example, a relative risk of 2 would imply that the predicted incidence is twice a high as would be expected given the area mean and local covariate values.
    \item The incidence rate ratio relative to some previous time point for spatio-temporal analyses. For example, we may be interested in comparing incidence to seven days prior; a value of 1.5 would imply that the incidence has risen 50\% in a week.
\end{enumerate}
We can extract any or all of these predictions using the function \texttt{extract\_preds()}, which will extract the predictions from the last model fit. For this example, we will extract from the crime data analysis the relative risk and incidence rate ratio. For region data models incidence predictions and incidence rate ratio are added to the region data, with the relative risk added to the grid data. 
\begin{CodeChunk}
\begin{CodeInput}
g2$extract_preds(type = c("rr","irr"),
                 irr.lag = 11,
                 popdens = "pop")
\end{CodeInput}
\end{CodeChunk}

\subsubsection{Hotspots}
The concept of a ``hotspot'' is often poorly or vaguely defined and it can mean different things to different people \citep{Lessler2017}. Generally, we can define it as an area where there is a high probability that some epidemiological measure(s) exceed predefined threshold(s). Here, we can use any of the outputs of the model listed above (incidence, relative risk, or incidence rate ratio). The \texttt{hotspots()} member function allows us to define a hotspot based on any combination of these criteria (connected logically by ``and''). For example, we can calculate the probability that the relative risk exceeds 1.5 across the grid:
\begin{CodeChunk}
\begin{CodeInput}
g2$hotspots(rr.threshold = 1.5, popdens = "pop", col_label = "rr")
\end{CodeInput}
\end{CodeChunk}
will add a column to the grid data labelled \code{rr} with the probabilities that the cell has a relative risk greater than 1.5. For region data models we cannot combing relative risk with the other statistics as the relative risk is predicted at the grid level, whereas the others are at the region level. For non-region models we could combine them, for example
\begin{CodeChunk}
\begin{CodeInput}
grid_data$hotspots(irr.threshold = 1.5,
                   irr.lag = 1,
                   rr.threshold = 2,
                   col_label="high_irr_and_rr")
\end{CodeInput}
\end{CodeChunk}
would return the probability that the incidence rate ratio relative to one month prior is greater than 1.5 and that the relative risk is greater than 2 and name the column in the data \texttt{high\_irr\_and\_rr}.

\subsection{Aggregating to other geographies}
It may be desirable to summarise the results of the analysis on a more familiar geography or to aggregate grid data to the region data. For example, we may wish to aggregate the results onto administrative areas that might be used to direct public health responses. In this example, we aggregate the relative risk results back onto the MSOA geography for the city. As with the covariate averaging we can either average with respect to the area or the population, here we’ll average with respect to the area.
\begin{CodeChunk}
\begin{CodeInput}
g2$region_data <- g2$aggregate_output(g2$region_data, zcols=c("rr"))
\end{CodeInput}
\end{CodeChunk}
The second row of Figure \ref{fig:bham_example} shows an example.

\subsection{Visualisation}
As the output of the analysis is an \texttt{sf} object (stored in the slot \texttt{g2\$grid\_data}, there are many different tools we can use to plot and visualise the results. We can use the default plot function in the \texttt{grid} class, which calls \texttt{sf}'s plot function for a variable in the grid data. For example, for the relative risk:
\begin{CodeChunk}
\begin{CodeInput}
grid_data$plot("rr")
\end{CodeInput}
\end{CodeChunk}
Other plotting tools include \pkg{ggplot2} with the \code{geom_sf} function; \pkg{tmap} provides a range of tools for map plotting including allowing multiple panels to be plotted. The \texttt{mapview} package provides functionality to create interactive overlays of \texttt{sf} data on OpenStreetMap maps. 

\subsection{Model parameters}
We may also be interested in the posterior samples of particular model parameters, which are contained in the output of the model fit. We note two important considerations for interpreting these parameters: 
\begin{itemize}
    \item For HSGP model fitting, the coordinates are scaled to $[-1,1]$, so the length scale parameters will be equivalently scaled. Use the \code{scale\_conversion\_factor()} function to extract these factors.
    \item Spatially-varying parameters of the linear predictor ($\gamma$) will differ from a non-spatial model. The parameters on spatially varying covariates will give an indication of the direction of an effect, but their magnitude may have different interpretations \citep{Hodges2010,Reich2006}.
\end{itemize}

\section{Conclusions}
The \pkg{rts2} packages provides multiple different methods and approximations for fitting the LGCP. The motivation for this package was to support real-time disease surveillance efforts by providing predictions in a fraction of the time of a full LGCP model with no approximation, while maintaining the reliability and calibration of the quantification of uncertainty. We can confirm that the first objective is met: in a set of comparisons, models that took days to run can now be run in minutes or hours, depending on the desired level of approximation. Indeed, one of the advantages of the two approximations used in this package is the ability to `tune' the approximation: at the extreme the approximations are equal to the full model. However, further research is required to establish the quality of the predictions and calibration of uncertainty, especially when compared with other `fast' alternatives like INLA and kernel-based methods. We have also provided functionality for data manipulation and extraction from model fits to support adoption and use of these methods. Future work and updates will aim to develop methods for comparing the quality of predictions generated by these and related methods. 

While our aim with this package is to support `real-time disease surveillance' the methods are applicable to multiple areas where case data are used to identify `hotspots' to support public and policy responses. We used an example with crime data. Other areas include geology, ecology, social policy, and many more.


\bibliography{ref}

\appendix

\section{A minimal, reproducible example}
In this section we provide code for a simple, reproducible example using simulated data.
\begin{CodeChunk}
\begin{CodeInput}
#create a square boundary and some random data points with dates
b1 <- st_sf(st_sfc(st_polygon(list(cbind(c(0,3,3,0,0),c(0,0,3,3,0))))))
npoints <- 20
dp <- data.frame(y=3*sqrt(runif(npoints)),
                 x=3*sqrt(runif(npoints)),
                 date=paste0("2021-01-",sample(11:20,npoints,replace = TRUE)))


# create a coarse grid over the area
g1 <- grid$new(b1,0.5)

# create the points sf object
dp <- create_points(dp,pos_vars = c('y','x'),t_var='date')

#create a random covariate over the area to act as population density
cov1 <- grid$new(b1,0.8)
cov1$grid_data$cov <- runif(nrow(cov1$grid_data))

# map the population density to the grid
g1$add_covariates(cov1$grid_data,
                  zcols="cov",
                  verbose = FALSE)

#aggregate the points to the grid
g1$points_to_grid(dp, laglength=3)

#set priors
g1$priors <- list(
  prior_lscale=c(0,0.5),
  prior_var=c(0,0.5),
  prior_linpred_mean=c(0),
  prior_linpred_sd=c(5)
)

# run the model
res <- g1$lgcp_fit(popdens="cov",
                   verbose = FALSE)

# extract the predictions
g1$extract_preds(res,
                 type=c("pred","rr"),
                 popdens="cov")
\end{CodeInput}
\end{CodeChunk}
We can compare our simulated data with the predicted incidence and relative risk.

\end{document}